\documentstyle[12pt]{article}

\begin{document}

\title{} \author{}

\begin{titlepage}

\title{On pion correction to quark mass
in next-leading order  of mean-field
expansion in NJL model}

\author{
R.G. Jafarov\\ {Baku State University}
, Baku, Azerbaijan\\
(e-mail: jafarov@hotmail.com)}
\date{}

\maketitle
\begin{abstract}
The correction to quark mass are calculated in Nambu-Jona-Lasinio
model with 4-dimensional cutoff regularization and with
dimensional-analytical regularization  in the next-to-leading
order of mean-field expansion. The analytical calculations show
that  the pion correction to quark mass is equal to zero.
Comparing the results in both regularizations one can signify that
the zero value of the pion correction to quark mass is the
regularization-independent fact of Nambu-Jona-Lasinio model.
\end{abstract}

\end{titlepage}

\newpage

\section{Introduction}

The chiral-symmetrical Nambu-Jona-Lasinio (NJL) model [1] with the quark
content [2] is one of the most successful theoretical laboratory for
investigation of the phenomenon of the spontaneous breakdown of chiral
symmetry and for study of the light hadrons in the non-perturbative region
[3,4]. The NJL model was intensively investigated also at finite temperature
and density [5] and with various external fields [6].

The nonrenormalizability of the NJL model implies a suitable choice of
regularization. The most common regularizations for NJL model traditionally
entail a four-dimensional cutoff (FDC) regularization[3,7,8] or
three-dimensional momentum cutoff regularization[3,4]. Other regularization
schemes are also used for NJL model [9].

Scalar meson contributions in the chiral quark condensate in the framework
of the dimensional-analytical regularization (DAR) have been calculated in
[10,11] . These contributions for physical values of parameters were found
to be significant and should be taken into account in choice of the
parameters values. The improved fit of parameters has been carried out in $%
SU_V\left( 2\right) \times SU_A\left( 2\right) $ -- NJL model.

In [8] a systematical comparison of the dimensional-analytically regularized
NJL model with the NJL model with FDC regularization has been done. Apart
from the corrections to chiral condensate, the corrections to quark mass in
both regularizations were also calculated. The numerical calculations at two
characteristic values of condensate showed that the pion contributions to
quark mass in both regularizations are equal to zero. Also it was supposed
in [8] that an absence of the pion corrections to quark mass is a
regularization-independent fact of NJL model.

The present work, which essentially based on results of [8], is devoted to
analytical calculations of the pion correction to quark mass.

\section{\ NJL model with dimensional-analytical regularization and with
four-dimensional cutoff. Meson contributions in chiral condensate}

The model under consideration contains up and down quarks fields $\psi
\left( x\right) $, each with $n_{c}$ colors. The $SU_V\left( 2\right) \times
SU_A\left( 2\right) -$invariant Lagrangian of this model is:

$$
L=\overline{\psi }i\widehat{\partial }\psi +\frac g2\left[ \left( \overline{%
\psi }\psi \right) ^2+\left( \overline{\psi }i\gamma _5\tau ^a\psi \right)
^2\right] , 
$$
where $\tau ^a$ are Pauli matrices normalized by $tr\left( \tau ^a\tau
^b\right) =2\delta ^{ab}$.

For formulating of the mean-field expansion we use an iteration scheme of
the solution of Schwinger-Dyson equation with the fermion bilocal source,
which has been developed in [12].

The unique connected Green's function in the leading mean-field
approximation is a single-particle Green's function (quark propagator):

$$
S^{\left( 0\right) }=\left( m-\widehat{p}\right) ^{-1}, 
$$
where the dynamical quark mass $m$ is a solution of the gap equation:

\begin{equation}
\label{1}1=-8ign_c\int \frac{d\widetilde{q}}{m^2-q^2}, 
\end{equation}
where $\ d\widetilde{q}\equiv d^4q/\left( 2\pi \right) ^4$.

The solution of the gap equation (1) in DAR and FDC regularization leads to
[8]:

\begin{equation}
\label{2} 1=\kappa \Gamma \left( \xi \right) \left( \frac{ M^2}{m^2}\right)
^{1+\xi }, 
\end{equation}

\begin{equation}
\label{3}1=\kappa _\Lambda \left( 1-\frac 1x\log \left( 1+x\right) \right) , 
\end{equation}
respectively. Here, $\kappa _\Lambda =\frac{gn_c\Lambda ^2}{2\pi ^2}%
,\,\kappa =\frac{gn_cm^2}{2\pi ^2},\,x=\frac{\Lambda ^2}{m^2};\,\xi $%
\thinspace \ and $\Lambda $\ are regularizations parameters in DAR and in
FDC regularization, correspondingly.

Meson contribution to chiral condensate can be calculated in the
next-to-leading term of the mean-field expansion.

Ratio of the first iteration condensate $\chi ^{\left( 1\right) }\,$\ to
leading-approximation condensate $\chi ^{\left( 0\right) }$ in the pion
channel has the form [8,10]:

\begin{equation}
\label{4}r_\pi =-\frac{24ign_c}{1-8ign_cJ}\int d\widetilde{p}d\widetilde{q} 
\frac{m^2-p^2+2pq}{\left( m^2-p^2\right) ^2\left[ m^2-\left( p-q\right)
^2\right] }A_\pi \left( q\right) , 
\end{equation}
where

$$
J=\int d\widetilde{p}\frac{m^2+p^2}{\left( m^2-p^2\right) ^2}. 
$$

In (4) the pseudoscalar amplitude $A_\pi $ is [10]:

\begin{equation}
\label{5}A_\pi =\frac{ig}{1+L_p}, 
\end{equation}
where $L_p\left( p\right) =ig\int d\widetilde{q}trS^{\left( 0\right) }\left(
p+q\right) \gamma _5$ $S^{\left( 0\right) }\left( q\right) \gamma _5-$ a
pseudoscalar quark loop.

Making use of the gap equation (1) we obtain the following form for
amplitude $A_\pi \,$ in momentum space: 
\begin{equation}
\label{6}A_\pi =\frac 1{4n_cp^2I_0\left( p^2\right) }, 
\end{equation}
where

\begin{equation}
\label{7} I_0\left( p^2\right) =\int \frac{d\widetilde{q}}{\left( m^2-\left(
p+q\right) ^2\right) \left( m^2-q^2\right) }. 
\end{equation}

The calculation of the integral (7) in both regularizations ( in $DAR$ and
in\ $FDC$ regularization) leads to: 
\begin{equation}
\label{8}I_0^{DAR}\left( p^2\right) =\frac i{\left( 4\pi \right) ^2}\frac
\xi \kappa F\left( 1+\xi ,1;3/2;\frac{p^2}{4m^2}\right) , 
\end{equation}

$$
I_0^{FDC}\left( p^2\right) =\frac i{\left( 4\pi \right) ^2}\left[ \log
\left( 1+x\right) -\frac x{1+x}F\left( 1,1;3/2;\frac{p^2}{4m^2\left(
1+x\right) }\right) -\right. 
$$

\begin{equation}
\label{9}\left. -\frac{p^{2}}{6m^{2}(1+x)}F\left( 1,1;5/2;\frac{p^{2}}{%
4m^{2}\left( 1+x\right) }\right) +\frac{p^{2}}{6m^{2}}F\left( 1,1;5/2;\frac{%
p^{2}}{4m^{2}}\right) \right] . 
\end{equation}
Here $F(a,b;c;z)$ is the Gauss hypergeometric function.

The pseudoscalar amplitude $\ A_\pi ^{pole}$ according to the pole $\left(
p^2=0\right) $ representations of the formulas (8) and (9) and the gap
equation (2) in both regularizations ( at $n=3$) takes the following forms:

\begin{equation}
\label{10}\left( A_\pi ^{pole}\right) ^{DAR}=\frac 1{12p^2I_0^{DAR}\left(
0\right) }=-\frac{2igm^2}{\xi p^2}, 
\end{equation}

\begin{equation}
\label{11}\left( A_\pi ^{pole}\right) ^{FDC}=\frac 1{12p^2I_0^{FDC}\left(
0\right) }=-i\frac{4\pi ^2}{3\left( \log (1+x)-\frac x{1+x}\right) p^2}. 
\end{equation}

As a measure of quantum fluctuations of the chiral condensate in pion
channel, it is used the first iteration condensate to the
leading-approximation condensate in pole approximation of amplitude (10) in $%
DAR$ [10]:

\begin{equation}
\label{12}r_\pi ^{DAR}=\frac 1{8\xi }, 
\end{equation}

Using the pole approximation of amplitude (11) in calculation of the ratio $%
r_\pi ^{FDC}$in Euclidean momentum space in\ $FDC$ regularization, we obtain
for ratio of the first iteration condensate to the leading approximation
condensate:

\begin{equation}
\label{13}r_\pi ^{FDC}=-\frac{\log (1+x)}{8\left( \log (1+x)-\frac
x{1+x}\right) }, 
\end{equation}

\section{ Pion correction to quark mass}

In [8] it has been obtained the formula for the correction to quark mass in
next-to-leading approximation of the mean-field expansion. We'll use that
formula having the following form:

\begin{equation}
\label{14}\frac{\delta m_\pi }m\cong b_\pi ^{(1)}\left( m^2\right) -a_\pi
^{(1)}\left( m^2\right) , 
\end{equation}
where $a_\pi ^{(1)}$ and $b_\pi ^{(1)}$ are the first order mass functions.
These functions are defined by the following equations:

\begin{equation}
\label{15} p^2a_\pi ^{(1)}\left( p^2\right) =-3\int \frac{d\widetilde{q}}{%
m^2-\left( p-q\right) ^2}A_\pi ^{pole}(q), 
\end{equation}

\begin{equation}
\label{16} b_\pi ^{(1)}\left( p^2\right) =r_\pi -3\int \frac{d\widetilde{q}}{%
m^2-\left( p-q\right) ^2}A_\pi ^{pole}(q) 
\end{equation}

Using in (15) and (16) the leading singularity approximation for $\left(
A_\pi ^{pole}\right) ^{DAR}$ (10) and \ $\left( A_\pi ^{pole}\right) ^{FDC}$
(11) and calculating the integrals in $\ DAR$ and \ $FDC$ regularization
(taking also into consideration (14)), we obtain for pion corrections to
quark mass the following expressions: 
\begin{equation}
\label{17}\left( \frac{\delta m_\pi }m\right) ^{DAR}=r_\pi ^{DAR}-\frac
1{8\xi }, 
\end{equation}

\begin{equation}
\label{18} \left( \frac{\delta m_\pi }m\right) ^{FDC}=r_\pi ^{FDC}+\frac{%
\log (1+x)}{8\left( \log (1+x)-\frac x{1+x}\right) } 
\end{equation}

From (17) and (18) and according to (12) and (13) it follows that the pion
contribution to quark mass is equal to zero. It means that the zero value of
the pion correction to quark mass is independent from the regularization
choice in NJL model.

\end{document}